\documentclass{article}
\usepackage{graphicx}
\usepackage{cite}
\begin{document}

\title{\bf Accurate energy spectrum for the quantum Yang-Mills mechanics with nonlinear color oscillations}
\author{Pouria Pedram\thanks{Email: p.pedram@srbiau.ac.ir}
\\  {\small Department of Physics, Tehran Science and Research Branch, Islamic
        Azad University, Tehran, Iran}}

\maketitle \baselineskip 24pt
\begin{abstract}
Yang-Mills theory as the foundation for quantum chromodynamics is a
non-Abelian gauge theory with self-interactions between vector
particles. Here, we study the Yang-Mills Hamiltonian with nonlinear
color oscillations in the absence of external sources corresponding
to the group $SU(2)$. In the quantum domain, we diagonalize the
Hamiltonian using the optimized trigonometric basis expansion method
and find accurate energy eigenvalues and eigenfunctions for one and
two degrees of freedom. We also compare our results with the
semiclassical solutions.

\vspace{5mm} {\it PACS numbers:} 11.10.Lm, 03.65.Ge
\end{abstract}
\maketitle

\section{Introduction}
All important theories of modern physics (except gravitation) such
as quantum electrodynamics, Weinberg-Salam electroweak theory, the
standard model of particle physics, and the grand unified theories
are quantized versions of the Yang-Mills theory. After quantization,
the quanta of the fields are usually interpreted as particles. The
Yang-Mills theory is a very active field of research as a gauge
theory based on the $SU(N)$ group. These theories have many subtle
and interesting properties which are still under several
investigations from physical or mathematical point of views. In
physics, the Yang-Mills theory is important since the predictions of
the Standard Model agree with the experiments with amazing accuracy.
In the context of mathematics, the Yang-Mills theory inspires
various important ideas in algebra, analysis, and geometry. Indeed,
the quantum Yang-Mills theories can be considered as a limit of a
more fundamental string theory. Note that the mass gap problem which
is discovered by physicists from experiment is one of the unsolved
millennium problems \cite{Clay}.

There has been much attention in the literature for solving
classical Yang-Mills equations and obtaining  non-perturbative
effects related with the ground state of quantum chromodynamics
\cite{pol1,pol2}. In Ref.~\cite{mat1} the classical Yang-Mills
equations without external sources are studied to address the
problem of asymptotic states of the theory and the structure of the
vacuum. In classical domain, the solutions exhibit nonlinear
oscillations of the color degrees of freedom similar to the massive
nonlinear plane waves in classical electrodynamics \cite{mat2}.

Here, we study this problem in the quantum domain and find the
accurate energy eigenvalues and eigenfunctions of the Yang-Mills
Hamiltonian describing nonlinear color oscillations. This problem is
not exactly solvable and we should resort to a numerical method. The
numerical methods for solving the time-independent Schr\"{o}dinger
equations can be categorized into two groups. The first group is
based on matrix diagonalization by representing the wave function in
terms of a finite orthogonal set of basis functions. Then, the
desired eigenvalues and eigenfunctions can be obtained by the
diagonalization of a Hamiltonian matrix computed from these basis
functions. The second group consists of the iterative methods that
are based on repeated numerical integrations of the Schr\"{o}dinger
equation with adjustments of the energy spectrum.

In this paper, following the first category, we use the optimized
trigonometric basis expansion method to find the solutions. Thus, we
expand the wave functions in terms of the trigonometric basis
functions obeying Dirichlet boundary condition. Then we use the
Rayleigh-Ritz variational scheme to optimize the domain of the basis
functions which leads to highly accurate solutions
\cite{taseli,Bhattacharyya,p0,p1,taseli2}.

\section{Yang-Mills equations without external sources}
Let us consider the Yang-Mills equations in the absence of
external sources corresponding to $SU(2)$ in the Minkowski space
\cite{mat1}
\begin{eqnarray}
\partial_\mu G_{\mu\nu}^{\,\,\,\,\,\,\,a}+g \epsilon^{abc} A_\mu^{\,\,\,b}
G_{\mu\nu}^{\,\,\,\,\,\,\,c}=0,\\
G_{\mu\nu}^{\,\,\,\,\,\,\,a}=\partial_\mu
A_\nu^{\,\,\,a}-\partial_\nu A_\mu^{\,\,\,a}+g \epsilon^{abc}
A_\mu^{\,\,\,b}A_\mu^{\,\,\,c},
\end{eqnarray}
where Latin indices range from 1 to 3, Greek indices range from 0 to
4, and $g$ is the gauge coupling constant. We look for a solution in
a coordinate system so that the Poynting vector
$T_{0j}=G_{0i}^{\,\,\,\,a}G_{ji}^{\,\,\,\,a}$ vanishes. Here
\begin{eqnarray}
T_{\mu\nu}=-G_{\mu\lambda}^{\,\,\,\,\,\,\,a}G_{\nu}^{\,\,\,\,\lambda
a}+\frac{1}{4}g_{\mu\nu}G_{\lambda\rho}^{\,\,\,\,\,\,a}G^{\lambda\rho
a},
\end{eqnarray}
is the energy-momentum tensor of the system. In the gauge
$\partial_i A_i^{\,\,\,a}=0$ and $A_0^{\,\,a}=0$ we have
\begin{eqnarray}\label{q1}
&&\epsilon^{abc} A_i^{\,\,b}\dot{A}_i^{\,\,c}=0,\\
&&\ddot{A}_i^{\,\,b}-\partial_j G_{ji}^{\,\,\,\,\,a}+g\epsilon^{abc}
A_j^{\,\,b}G_{ji}^{\,\,\,\,\,c}=0,\\
&&\dot{A}_i^{\,\,a}G_{ij}^{\,\,\,\,\,a}=0.\label{q3}
\end{eqnarray}
Using Eqs.~(\ref{q1}) and (\ref{q3}) we find
\begin{eqnarray}\label{w1}
\dot{A}_i^{\,\,a}\left(\partial_jA_i^{\,\,a}-\partial_iA_j^{\,\,a}\right)=0.
\end{eqnarray}
So the sufficient condition to satisfy Eq.~(\ref{w1}) is
\begin{eqnarray}
(a)\,\,\partial_jA_i^{\,\,a}=0,\hspace{.5cm}\hspace{.5cm}(b)\,\,\dot{A}_i^{\,\,a}=0,\hspace{.5cm}\hspace{.5cm}(c)\,\,
\partial_jA_i^{\,\,a}-\partial_iA_j^{\,\,a}=0.
\end{eqnarray}
In the following sections we consider case $(a)$ where $A_i^{\,\,a}$
have no spacial dependence.

\section{Color space with two degrees of freedom}
If we take $\partial_j A_i^{\,\,\,a}=0$ ($i,j=1,2,3$), the potential
depends only on the time and the Yang-Mills equations are given by
\cite{mat2}
\begin{eqnarray}
\ddot{A}_i^{\,\,\,a}+g^2\left(A_j^{\,\,\,b}A_j^{\,\,\,b}A_i^{\,\,\,a}-A_j^{\,\,\,a}A_j^{\,\,\,b}A_i^{\,\,\,b}\right)=0,
\end{eqnarray}
where dot denotes differentiation with respect to the time. The
Hamiltonian for this system is
\begin{eqnarray}\label{HYM}
H=\frac{1}{2}(\dot{A}_i^{\,\,\,a})^2+\frac{g^2}{4}\left[\left(A_i^{\,\,\,a}A_i^{\,\,\,a}\right)^2-\left(A_i^{\,\,\,a}A_j^{\,\,\,a}\right)^2\right].
\end{eqnarray}
Following Ref.~\cite{mat1}, one solution of Eq.~(\ref{HYM}) can be
expressed as the following nine-parameter form
\begin{eqnarray}
A_i^{\,\,\,a}=\frac{1}{g}O_i^{\,\,\,a}f^{(a)}(t),
\end{eqnarray}
where there is no summation over $a$, $O_i^{\,\,\,a}$ is a constant
orthogonal matrix i.e. $O_i^{\,\,\,a}O_i^{\,\,\,b}=\delta^{ab}$, and
$f^{(a)}$ denotes the three colors. Now we find
\begin{eqnarray}\label{eqf}
\ddot{f}^{(a)}+f^{(a)}\left(\mathbf{f}^2-f^{(a)2}\right)=0,
\end{eqnarray}
where $\mathbf{f}^2=f^{(1)2}+f^{(2)2}+f^{(3)2}$. For the case
$f^{(3)}=0$, if we introduce $x=f^{(1)}$ and $y=f^{(2)}$, the
nonlinear equations of motion are
\begin{eqnarray}
\ddot{x}+xy^2=0,\hspace{2cm}\ddot{y}+yx^2=0.
\end{eqnarray}
So, the corresponding Hamiltonian is given by
\begin{eqnarray}\label{eqH}
H=\frac{1}{2}p_x^2+\frac{1}{2}p_y^2+\frac{1}{2}x^2y^2,
\end{eqnarray}
which is well-known in the classical and quantum chaos studies
\cite{r0,r1,r2,r3,r4,r5}. The quartic potential $V(x,y)=x^2y^2$
appears in various branches of science such as chemistry,
astrophysics, and cosmology. In particular, in the classical scalar
electrodynamics without self-interaction of the scalar field, when
only a single component of the electromagnetic gauge field is
nonvanishing, it is classically equivalent to the two-dimensional
dynamical system with the potential $x^2y^2$ that shows a strong
chaotic behavior \cite{sed}. This potential also appears in the
homogeneous limit of the Yang-Mills equations \cite{QCD,15}.

\section{The optimized trigonometric basis-set expansion method}
In the quantum domain, the Hamiltonian (\ref{eqH}) leads to the
following two-dimensional time-independent Schr\"{o}dinger equation
($\hbar=1$)
\begin{equation}\label{ODE}
-\left(\frac{\partial^2}{\partial x^2}+\frac{\partial^2}{\partial
y^2}\right)\psi(x,y)+V(x,y)\psi(x,y)=\epsilon\,\psi(x,y),
\end{equation}
where $V(x,y)=x^2y^2$, $\epsilon=2E$, $E$ is the energy of the
system, i.e. $H\psi(x,y)=E\psi(x,y)$, and the wave function
$\psi(x,y)$ usually satisfies the following condition
\begin{equation}\label{boun1}
\lim_{x,y\rightarrow\infty}\psi(x,y)=0.
\end{equation}
To find the approximate energy eigenvalues, we implement a
two-dimensional Rayleigh-Ritz variational method whereas the
solutions are determined upon two independent variables, namely the
truncated domains in $x$ and $y$ directions. The important point of
the method is the consideration of a truncated domain of the
independent variables $x$ and $y$ so that
\begin{equation}
x\in\left[-\frac{L_x}{2},+\frac{L_x}{2}\right],\hspace{2cm}y\in\left[-\frac{L_y}{2},+\frac{L_y}{2}\right],
\end{equation}
and the modification of the usual boundary condition (\ref{boun1}).
Thus, the problem is finding the solution of $H \psi = E\psi$
subject to the Dirichlet boundary conditions
\begin{equation}
\psi\left(-\frac{L_x}{2},y\right)=\psi\left(+\frac{L_x}{2},y\right)=\psi\left(x,-\frac{L_y}{2}\right)=\psi\left(x,+\frac{L_y}{2}\right)=0,
\end{equation}
for all values of $x$ and $y$ on the boundaries of the finite
rectangular region. This method gives highly accurate results if
both the truncated domain and the number of the basis functions are
adjusted properly in one \cite{taseli,Bhattacharyya,p1} or two
\cite{taseli2} dimensions.

To proceed further, we choose a finite set of the trigonometric
basis functions obeying Dirichlet boundary condition. Moreover, to
simplify the diagonalization procedure, we shift the domain to
$0<x<L_x$ and $0<y<L_y$ and write
\begin{eqnarray}
\psi(x,y)=\sum_{m,l=1}^{\infty} A_{ml}
 \sin\left(\frac{m \pi x}{L_x}\right)\,\sin\left(\frac{l \pi
y}{L_y}\right). \label{eqpsitrigonometric}
\end{eqnarray}
Now since we can write $ V(x,y) \psi(x,y)=\sum_{m,l} B_{ml}
\sin\left(\frac{m \pi x}{L_x}\right)\sin\left(\frac{l \pi
y}{L_y}\right),\label{eqV} $ we find
\begin{eqnarray}
\left[\left(\frac{m \pi}{L_x}\right)^2+\left(\frac{l
\pi}{L_y}\right)^2\right] A_{ml}+B_{ml}=\epsilon\,
A_{ml}.\label{eqAB2}
\end{eqnarray}
The matrix $B$ is determined by $ B_{m,l}= \sum_{m',l'}
C_{mm'll'}A_{m'l'} $ where
\begin{eqnarray}\label{cmm}
C_{mm'll'}=\frac{4}{L_xL_y}\int_{0}^{L_x}\int_{0}^{L_y}
\sin\left(\frac{m \pi x}{L_x}\right)\sin\left(\frac{l \pi
y}{L_y}\right)V(x,y)\sin\left(\frac{m' \pi
x}{L_x}\right)\sin\left(\frac{l' \pi y}{L_y}\right) dxdy.
\end{eqnarray}
Therefore,  Eq.~(\ref{eqAB2}) can be rewritten as
\begin{eqnarray}
\left[\left(\frac{m \pi}{L_x}\right)^2+\left(\frac{l
\pi}{L_y}\right)^2\right] A_{ml}+ \sum_{m',l'} C_{mm'll'}
A_{m'l'}=\epsilon A_{ml}.\label{eqAC}
\end{eqnarray}
Notice that the presence of the potential term leads to nonzero
coefficients $C_{m,m',l,l'}$ in Eq.~(\ref{eqAC}), which couples all
of the matrix elements of $A$. To diagonalize the Hamiltonian, we
select the first $N^2$ basis functions by letting the indices $m$
and $l$ run from 1 to $N$. For this case, we replace
Eq.~(\ref{eqpsitrigonometric}) with the expansion of the solutions
in terms of $N\times N$ basis functions in two-dimensional space,
namely $\psi(x,y)=\sum_{m,l=1}^{N} A_{ml}
 \sin\left(\frac{m \pi x}{L_x}\right)\,\sin\left(\frac{l \pi
y}{L_y}\right)$. Then we replace the square matrix $A$ with a column
vector $\tilde A$ with $N^2$ elements, so that any element of $A$
corresponds to one element of $\tilde A$. With this replacement
Eq.~(\ref{eqAC}) can be written as
\begin{eqnarray}
D\tilde A=\epsilon\tilde A , \label{eqmatrix}
\end{eqnarray}
where $D$ is a square matrix with $N^2 \times N^2$ elements and can
be obtained from Eq.~(\ref{eqAC}). So the solution to this matrix
equation simultaneously yields $N^2$ sought after energy eigenvalues
and eigenstates, namely $\psi_n(x,y)$ where $n=1,2,\ldots,N^2$. Now
the optimization procedure is the adjustment of $L_x$ and $L_y$ for
each $N$ and we denote these optimal quantities by $\hat{L_x}(N)$
and $\hat{L_y}(N)$ which correspond to the minimum value of
$\epsilon(N,L_x,L_y)$ for a fixed $N$. Indeed, highly accurate
solutions can be obtained upon using these optimal lengths.

For our case, since the potential contains the interchange symmetry,
i.e., $V(x, y) = V(y, x)$ we solve the problem on a square domain
putting $L_x = L_y$ \cite{taseli2}. Note that we only imposed the
condition of the vanishing of the wave functions at the boundaries.
Physically, it means that the potential is infinite outside. Indeed,
the potential is $V(x,y)=x^2y^2$ inside and $V(x,y)=\infty$ outside.
This situation is similar to the particle in a box where the
potential inside does not vanish anymore. This approximation is
valid for the low-lying energy eigenvalues where their corresponding
eigenstates almost vanish at the boundaries.

\begin{table}
  \centering
\begin{tabular}{|c|c|c|c|} \hline
$(m,l)$& $n$             & $\epsilon_n$            & error                                    \\ \hline
 (1,1) &      1          &    1.10822315780256     & $1.19\times10^{-10}$                     \\ \hline
 (1,2) &      2          &    2.37863785124994     & $1.16\times10^{-8}$                      \\ \hline
 (2,1) &      3          &    2.37863785124996     & $1.16\times10^{-8}$                      \\ \hline
 (2,2) &      4          &    3.05608156130323     & $2.06\times10^{-7}$                      \\ \hline
 (3,3) &      5          &    3.51495134040797     & $1.12\times10^{-6}$                      \\ \hline
 (2,3) &      6          &    4.09348955687600     & $9.38\times10^{-6}$                      \\ \hline
 (3,2) &      7          &    4.09348955687604     & $9.38\times10^{-6}$                      \\ \hline
 (4,4) &      8          &    4.75298944936096     & $9.32\times10^{-5}$                      \\ \hline
 (5,5) &      9          &    4.98538290136962     & $1.75\times10^{-5}$                      \\ \hline
 (6,6) &      10         &    5.01127928161308     & $4.59\times10^{-11}$                     \\ \hline
 (1,3) &      11         &    5.50103621623983     & $7.92\times10^{-4}$                      \\ \hline
 (3,1) &      12         &    5.50103621623990     & $7.92\times10^{-4}$                      \\ \hline\hline
(11,11)&      21         &    8.07437393671447     & $6.64\times10^{-9}$                      \\ \hline
(14,14)&      26         &    9.27305945794927     & $3.36\times10^{-8}$                      \\ \hline
(18,18)&      34         &    11.4718771513251     & $7.24\times10^{-7}$                      \\ \hline
(23,23)&      45         &    13.8662683175987     & $8.33\times10^{-6}$                      \\ \hline
       &   $\hat{L}(42)$ &                                        \multicolumn{2}{c|}{$15.53$}\\ \hline
\end{tabular}
  \caption{The first 12 states and some
highly exited states with $N=42$ basis functions.} \label{TableQCD}
\end{table}
\begin{figure}
\centering
\includegraphics[width=10cm]{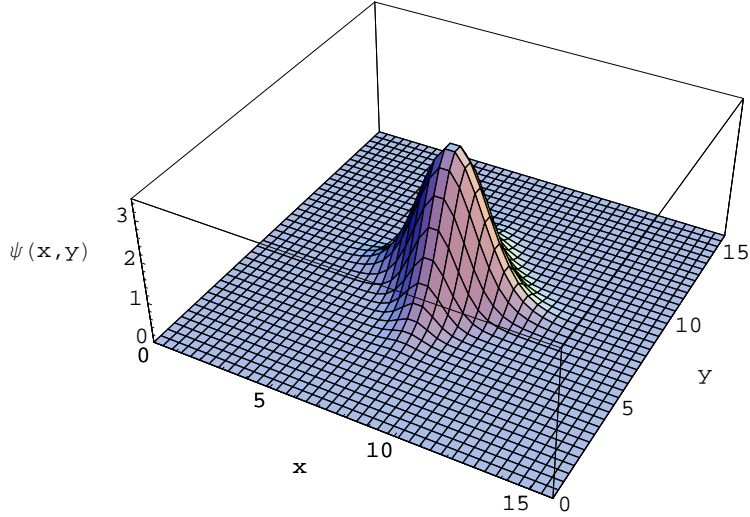}
\caption{The ground state wave function using $N=42$ and
$\hat{L}(42)=15.53$. } \label{figQCD1}
\end{figure}

\begin{figure}
\centerline{\begin{tabular}{ccc}
\includegraphics[width=8cm]{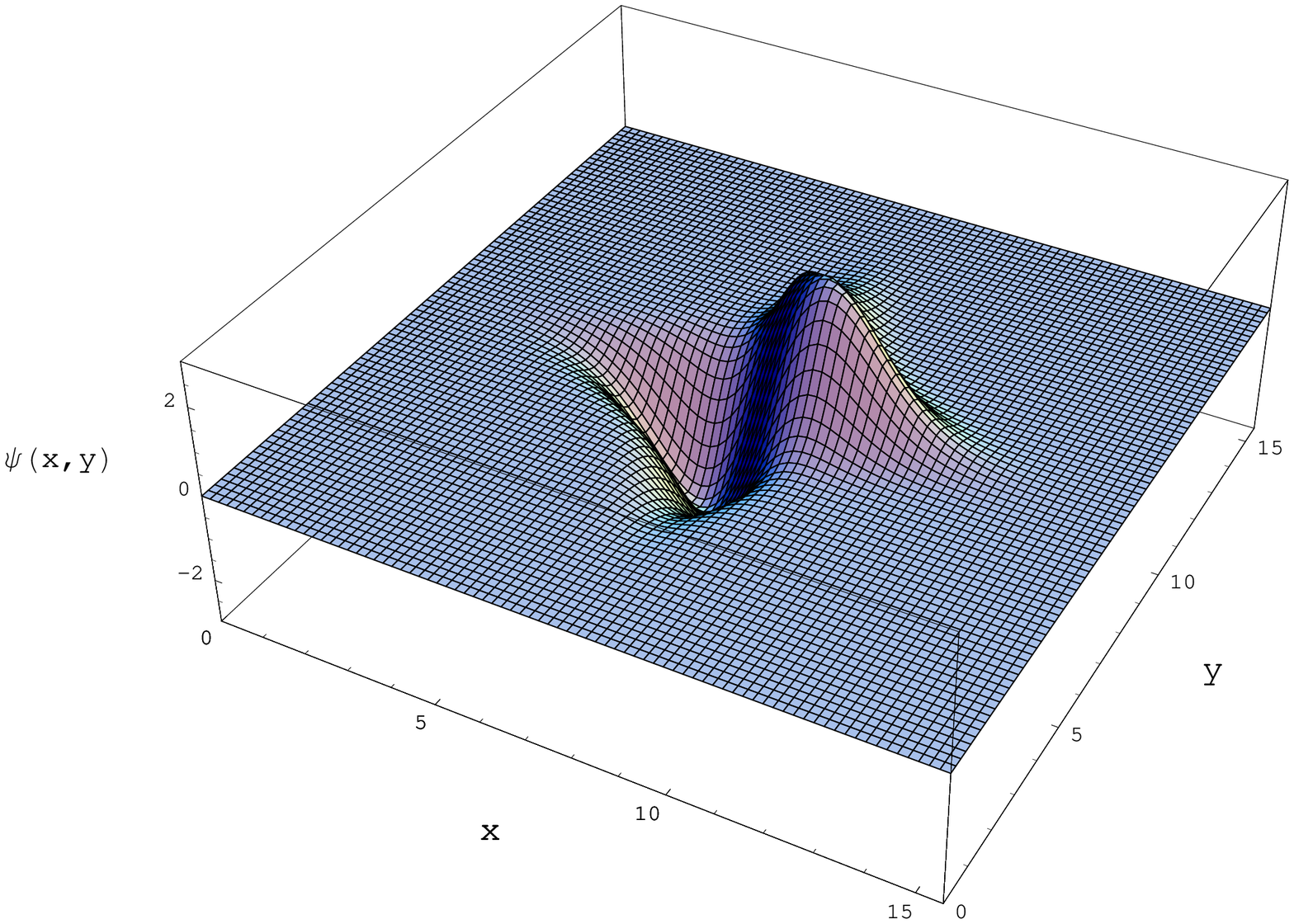}
 &\hspace{2.cm}&
\includegraphics[width=8cm]{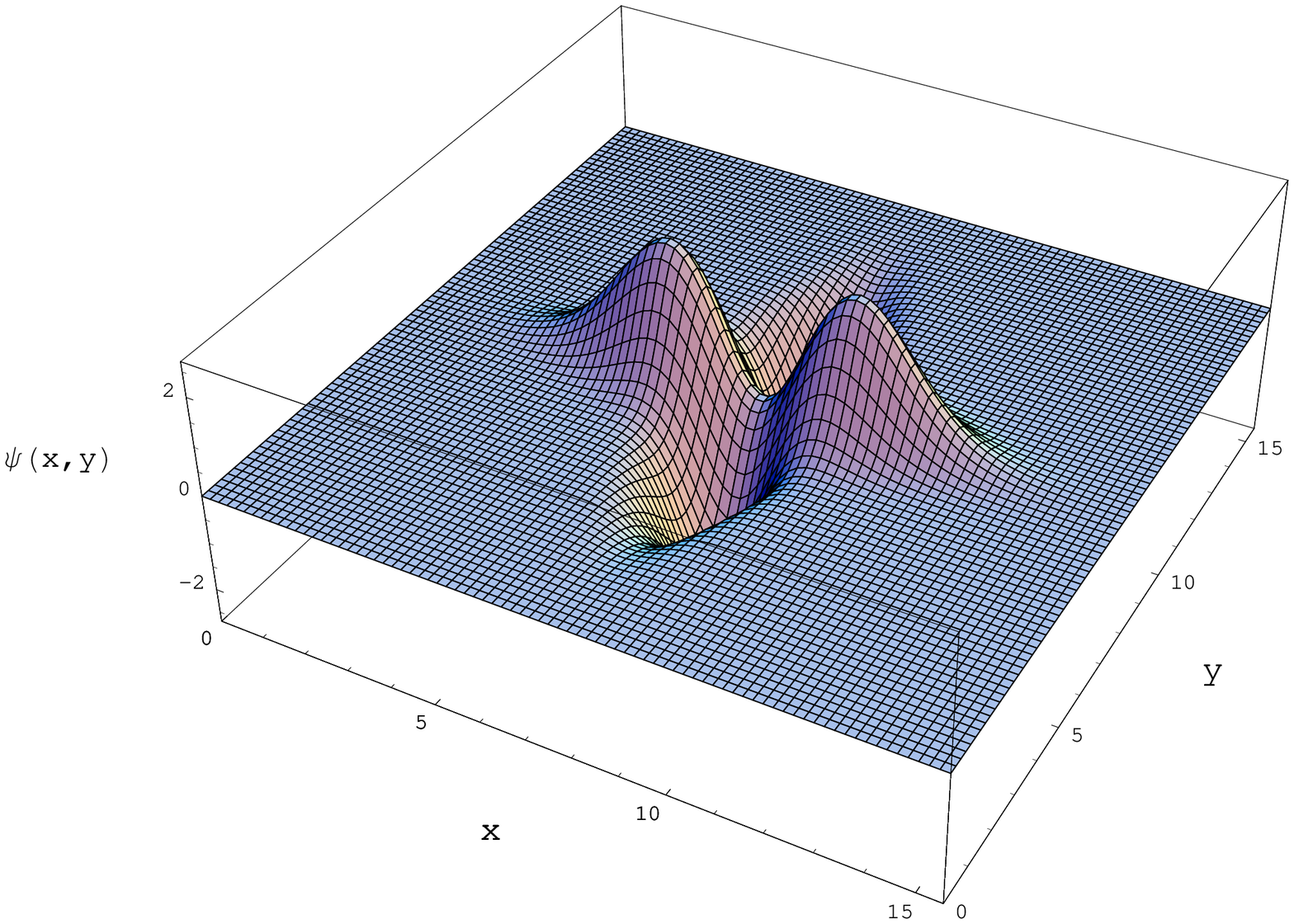}\\
\includegraphics[width=8cm]{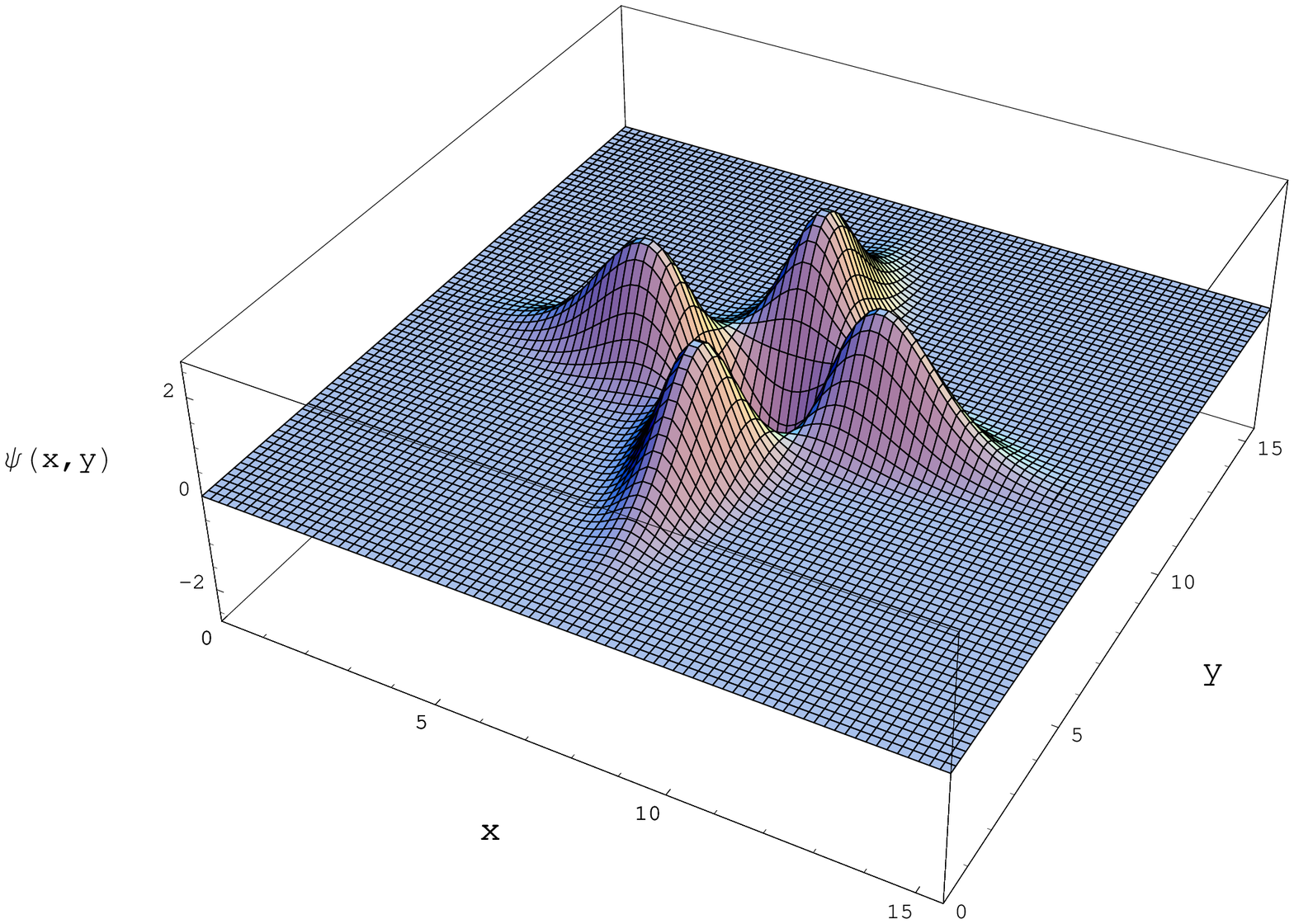}
 &\hspace{2.cm}&
\includegraphics[width=8cm]{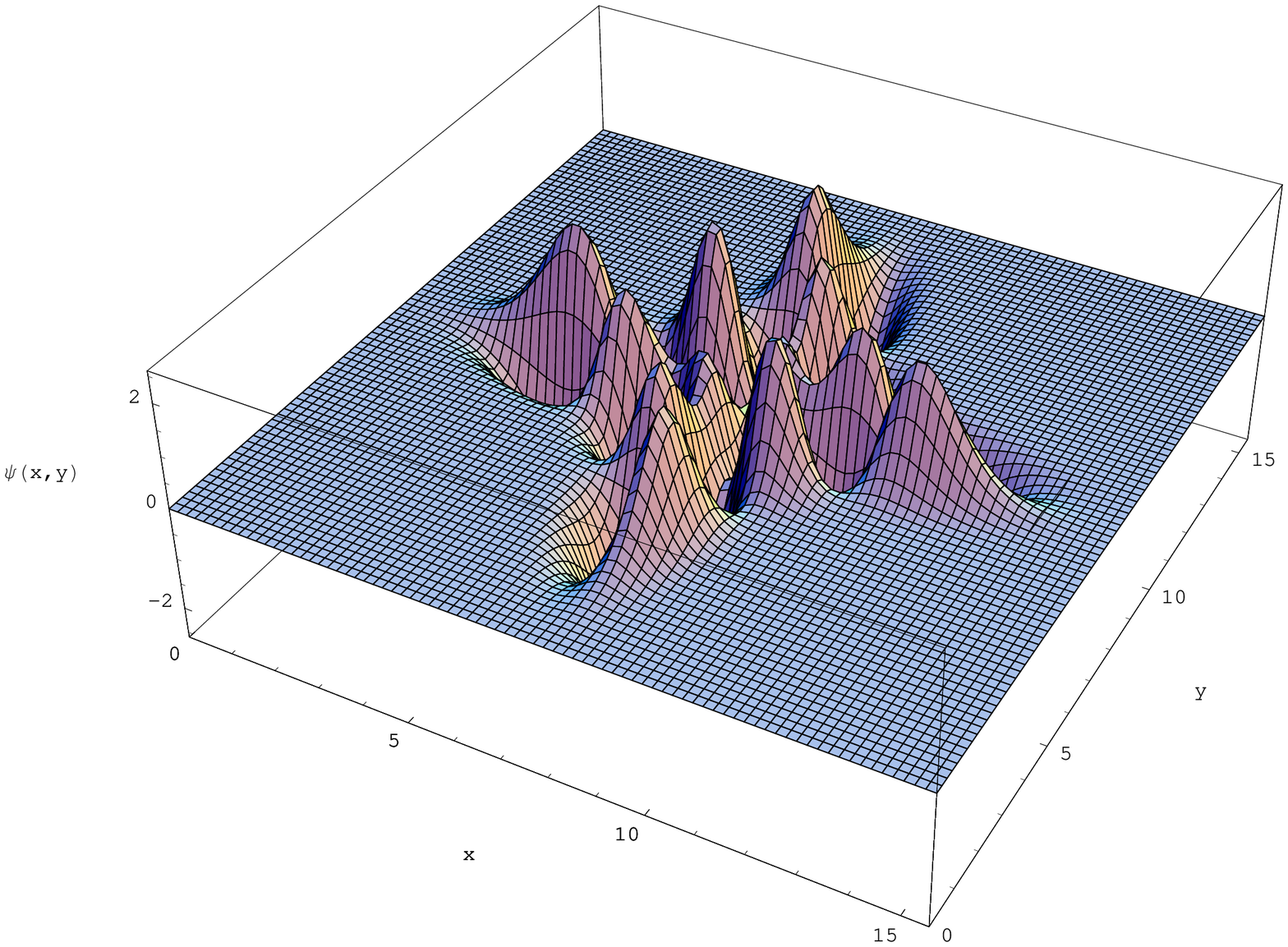}
\end{tabular}}
\caption{Upper left: 2nd state, upper right: 4th state, lower left:
5th state, and lower right: 44th state.} \label{figQCD2}
\end{figure}

In Table \ref{TableQCD}, we have shown the eigenvalues for the first
12 states and some highly exited ones, namely, $n=\{21,26,34,45\}$.
These highly exited states are chosen for their high accuracy due to
their symmetric form, and the fact that we have chosen
$\hat{L}_x(N)=\hat{L}_y(N)=\hat{L}(N)$. Figure \ref{figQCD1} shows
the ground state wave function. Notice the slight over extension of
the wave function in the $x$ and $y$ directions due to the
particular form of the potential.  In Fig.~\ref{figQCD2} we have
shown the wave functions for the second, forth, fifth, and forty
forth eigenstates. The third eigenstate is degenerate with the
second state and can be obtained from it by 90 degree rotation. Note
that, as it can be seen from Eqs.~(\ref{cmm}) and (\ref{eqAC}),
exchanging quantum numbers $m$ and $l$ leads to the same eigenvalue
equation. Thus, energy levels with exchanged quantum numbers should
be degenerate. In particular, the values of $(m,l)$ which are shown
in Table \ref{TableQCD} explicitly exhibit the degeneracy of the
problem.

\section{Color space with one degree of freedom}
In this section we are interested to study the particular solution
of Eq.~(\ref{eqf}), namely $f^{(1)}=f^{(2)}=f^{(3)}=f$ \cite{mat2}.
For this case the equation of motion is
\begin{eqnarray}
\ddot{f}+2f^3=0.
\end{eqnarray}
So the solution is
\begin{eqnarray}
f(t)=\left(\frac{2g^2}{3}\right)^{1/4}\mu\,
cn\left[\left(\frac{8g^2}{3}\right)^{1/4}\mu(t+t_0);\frac{1}{\sqrt{2}}\right],
\end{eqnarray}
where $cn(x; k)$ is the Jacobian elliptic cosine of argument $x$ and
modulus $k$, $\mu^4$ is the energy density $T_{00}$ in the present
coordinate system, and $t_0$ is the arbitrary origin of the time. So
the solution is periodic with period
$T=\left(3/8g^2\right)^{1/4}(4/\mu)K(1/\sqrt{2})$ where $K(x)$ is
the complete elliptic integral of the first kind. Now, if we take
$x=f$ the corresponding Hamiltonian reads
\begin{eqnarray}\label{anh}
H=\frac{1}{2}p_x^2+\frac{1}{2}x^4.
\end{eqnarray}
Note that the quartic potential has attracted much attention in literature because of its
similarity to the $\lambda \phi^4$ quantum field theory as the prototype of spontaneous
symmetry breaking.

In the quantum domain the above Hamiltonian results in the following
one-dimensional Schr\"odinger equation
\begin{eqnarray}
-\frac{\partial^2\psi(x)}{\partial
x^2}+x^4\psi(x)=\epsilon\,\psi(x),
\end{eqnarray}
where $\epsilon=2E$. This equation is the Schr\"odinger equation for
the anharmonic oscillator in one-dimension and is not exactly
solvable. However, we can use the optimized trigonometric basis-set
expansion method to find the highly accurate solutions
\cite{taseli,Bhattacharyya,p0,p1}. On the other hand, the
semiclassical approximation ($\hbar=1$)
\begin{eqnarray}
\oint p_x\,
\mathrm{d}x=2\pi\left(n-\frac{1}{2}\right),\hspace{2cm}n=1,2,\ldots,
\end{eqnarray}
results in
\begin{eqnarray}
E_n=\frac{1}{2}\left[\frac{\sqrt{\pi}\,\Gamma(7/4)}{\Gamma(5/4)}\left(n-\frac{1}{2}\right)\right]^{4/3}.
\end{eqnarray}

\begin{table}
  \centering
\begin{tabular}{|c|c|c|} \hline
$n$&$\epsilon_n^{\mathrm{exact}}$&  WKB                         \\ \hline
   1         &    0.530181045    & 0.4335                       \\ \hline
   3         &    3.727848969    & 3.7069                       \\ \hline
   5         &    8.130913009    & 8.1168                       \\ \hline
   7         &    13.26423559    & 13.253                       \\ \hline
   9         &    18.96150051    & 18.952                       \\ \hline
   11        &    25.12812726    & 25.120                       \\ \hline
   13        &    31.70152349    & 31.694                       \\ \hline
   15        &    38.63660024    & 38.630                       \\ \hline
   17        &    45.89903340    & 45.893                       \\ \hline
   19        &    53.46165369    & 53.456                       \\ \hline
   21        &    61.30231950    & 61.297                       \\ \hline
\end{tabular}
  \caption{The low-lying energy eigenvalues for the one-dimensional anharmonic oscillator (\ref{anh}) using
  highly accurate numerical and semiclassical schemes.} \label{Tab2}
\end{table}

In Table \ref{Tab2} we have reported the low-lying energy
eigenvalues for the anharmonic oscillator using the optimized
trigonometric basis-set expansion and the semiclassical methods. As
it can be seen from the table the agreement between the exact and
semiclassical results increases for the excited states with large
quantum numbers.

\section{Conclusions}
Since the Yang-Mills equations are nonlinear, they are not
explicitly solvable in general. This is similar to the Einstein
equations for the gravitational field but unlike the Maxwell
equations for the electromagnetic field.  However, like the Maxwell
equations they describe  massless waves that travel at the speed of
light at the classical level. Although the classical non-abelian
gauge theory is within the reach of established mathematical
methods, the precise definition of quantum gauge theory in four
dimensions is still unclear. In this paper and in the context of the
first quantization, we have studied the quantum Yang-Mills mechanics
in the absence of external sources corresponding to the group
$SU(2)$. In the classical domain, this system in a coordinate system
in which the Poynting vector vanishes, exhibited nonlinear
oscillations of the color degrees of freedom. We solved its
corresponding Schr\"{o}dinger equations in quantum domain in which
the potential term becomes $x^4$ and $x^2y^2$ in one and two
dimensions, respectively. Using the optimized trigonometric basis
expansion method we found the accurate energy eigenvalues and
eigenstates and compared the results with the semiclassical
solutions.


\end{document}